\documentclass[twocolumn]{aastex631}

\usepackage{newtxtext,newtxmath}
\usepackage[T1]{fontenc}
\usepackage{hyperref}
\usepackage{graphicx}	
\usepackage{amsmath}	
\usepackage{gensymb}
\usepackage{soul}

\shorttitle{Migration via Massloss}
\shortauthors{Hanf et al.}

\begin{document}

\title[Migration via Mass Loss]{Orbital Migration through Atmospheric Mass Loss}

\author{Benjamin Hanf}
\altaffiliation{Both authors contributed equally to this manuscript}
\affiliation{Pomona College,
333 N College Way, 
Claremont, CA 91711}


\author{Will Kincaid}
\altaffiliation{Both authors contributed equally to this manuscript}
\affiliation{Harvey Mudd College, 
301 Platt Blvd.
Claremont, CA 91711}

\author{Hilke Schlichting}
\affiliation{UCLA, 
595 Charles E. Young Drive East, 
Los Angeles, CA 90095}

\author{Livan Cappiello}
\affiliation{Harvey Mudd College, 
301 Platt Blvd.
Claremont, CA 91711}

\author{Daniel Tamayo}
\affiliation{Harvey Mudd College, 
301 Platt Blvd.
Claremont, CA 91711}
 
\begin{abstract}
Atmospheric mass loss is thought to have strongly shaped the sample of close-in exoplanets. 
These atmospheres should be lost isotropically, leading to no net migration on the planetary orbit.
However, strong stellar winds can funnel the escaping atmosphere into a tail trailing the planet.
We derive a simple kinematic model of the gravitational interaction between the planet and this anisotropic wind, and derive expressions for the expected migration of the planet. Over the expected range of parameters, we find typical migrations of a few tenths to a few percent inward. We argue that this modest migration may be observable for planet pairs near mean motion resonances, which would provide an independent observational constraint on atmospheric mass loss models.

\end{abstract}

\keywords{Exoplanet Migration (2205)}

\section{Introduction}

A central goal of exoplanet science is to infer the composition of planets beyond our solar system. This provides important constraints on theories of planet formation and our understanding of habitability on other worlds.
Unfortunately, one cannot extract planetary compositions uniquely from only measurements of mass and radius \citep[e.g.,][]{Neil20}. 
Significant effort has thus gone into uncovering and understanding observational trends which might help break these degeneracies.

A particularly influential discovery has been that the radius distribution of planets smaller than Neptune is bimodal, with a valley separating planets $\lesssim 1.8 R_\oplus$ from a second pileup with bodies of $2-3 R_\oplus$ \citep[see Fig. 7 in][]{Fulton17, vanEylen18}. Most compositional investigations agree that the small-planet population represents bare, rocky cores \citep[e.g.,][]{Owen17, Zeng19, Gupta19}, while the typical interpretation has been that the larger $2-3 R_\oplus$ planets require volatile H/He envelopes with mass fractions of order $1\%$ to explain their low measured masses \citep{Wolfgang15, Rogers15}.

Such a radius dichotomy was naturally predicted by atmospheric mass loss models, either through photoevaporation \citep{Owen13, Lopez13}, or through leftover heat from accretion a.k.a. core-powered mass loss \citep{GSS16, GSS18, Gupta19}.
These models predict that all close-in planets start with significant H/He atmospheres that are lost over time. 
Mass-loss from more massive planets stabilizes at the observed radius pileup of $2-3 R_\oplus$, at which they are most resistant to erosion, while lower-mass planets lose their envelopes entirely to become bare cores \citep{Owen17}. In this picture, a significant fraction of the short-period exoplanet population may have lost a few percent of its mass in the form of an escaping atmosphere.
This motivates investigating how atmospheric mass loss might affect these planets' orbits.

Previous studies have considered an anisotropic loss of atmospheric mass \citep{boue12, Teyssandier15, Fujita22} or photons \citep{Fabrycky8}, that, like a rocket, causes a change in momentum in the opposite direction. 
While they consider different physical mechanisms (the Yarkovsky effect, differential stellar heating), the material is preferentially ejected behind the planet, causing the planet to gain orbital angular momentum and move to a \textit{wider} semimajor axis. 
Such models typically predict up to a few percent outward migration \citep{Teyssandier15}; \cite{wang23} argue that larger-scale migration is ruled out in compact multiplanet systems, since this would lead to dynamical instabilities inconsistent with observed configurations.
However, such models agree that there should be no migration in the limit one would naïvely expect, where the atmosphere is lost isotropically \citep[e.g.,][]{Murray-Clay09, Wang21}.

\begin{figure}
    \centering \resizebox{\columnwidth}{!}{\includegraphics{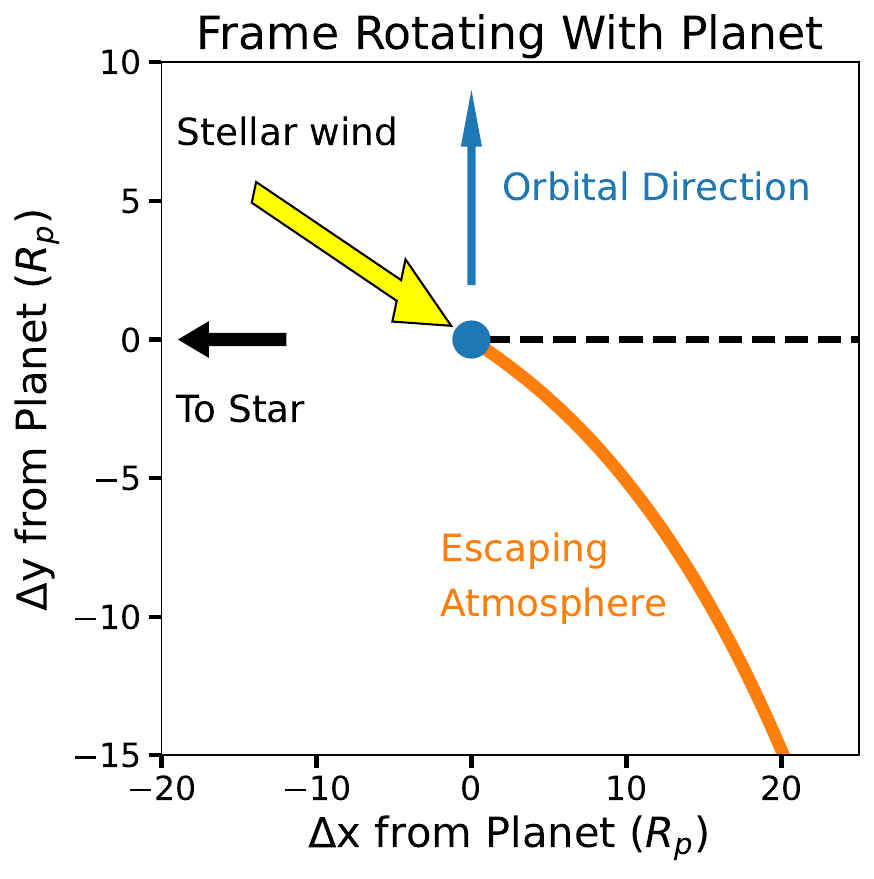}}
    \caption{Frame co-rotating with the planet (blue circle), with the orbital direction upward. The star is off the frame along the negative x axis. Due to the planet's orbital speed, the approximately radial stellar wind has a head-on component in the planet's frame \citep[see also Fig. 1 of][]{Wang21}. The escaping atmosphere (orange) is funneled behind the planet by the stellar wind.
    \label{fig:diagram}}
\end{figure}

Nevertheless, numerical studies have shown that sufficiently strong stellar winds (yellow arrow in Fig.\:\ref{fig:diagram}) can redirect an initially isotropic atmospheric outflow, funneling it into a tail (orange line in Fig.\:\ref{fig:diagram}) that trails the planet \citep{Wang18, McCann19, MacLeod22, Wang21b}.
In this scenario, we will show that the initially isotropic outflow generates no net migration, but the resulting atmospheric tail gravitationally pulls the planet backward, causing it to lose orbital energy.  In contrast to previous studies, we predict this mechanism should cause inward migration, which in particular can lead to different observational signatures near mean motion resonances (see Sec.\:\ref{sec:jumps}).

Simulations modeling the hydrodynamics of this process, incorporating radiative transfer and chemistry, are computationally expensive \citep[e.g.,][]{Wang18} 
and thus limited to integrating only a few dozen orbital periods. Hydrodynamic simulations are therefore not able to probe orbital evolution on the timescales over which the atmosphere is expected to be lost, $\sim$100 Myr to $\sim$1 Gyr \citep{Owen17, guptaschlichting2020, Berger_2020}.
However, given that the flow becomes supersonic shortly after launch and its total energy is thus dominated by its kinetic component \citep{Murray-Clay09}, much cheaper N-body integrations of appropriately launched ballistic particles give a good approximation for how the flow of escaping gas is shaped by the star's tidal gravitational field near the planet \citep{McCann19}.
In this paper we therefore develop an approximate analytical model, and compare it to computationally accessible N-body integrations.

In Sec.\:\ref{sec:model}, we derive an expression for the change in period, beginning with the simplest case, and later adding parameters to more accurately model atmospheric escape. We then analyze the limiting case of total atmospheric stripping, and test the model against N-body simulations. In Sec.\:\ref{sec:results} we develop a fiducial model for the change in period, as well as an upper and lower bound for the magnitude of migration assuming plausible physical parameters. Finally, we discuss the observational consequences of such orbital migration and conclude in Sec.\:\ref{sec:conclusion}.

\section{Analytical Model} \label{sec:model}

\subsection{Impulse Approximation} \label{sec:impulse}

The gravitational interaction between an escaping atmosphere and the planet decreases with separation distance.
The characteristic timescale for the interaction is thus set by the time it takes for the atmosphere to travel roughly a distance $R_p$, where $R_p$ is the radius of the planet, doubling its separation.
The distribution of speeds for escaping particles typically decays toward higher velocities, so we assume that most of the escaping atmosphere is launched near the escape speed \footnote{For example, in the hydrodynamic simulations of an evaporating 5 $M_{\oplus}$ planet by \cite{Wang18}, the gas escapes with a residual speed at large separations of only $\sim 1\%$ of the planet's escape speed.} $v\sim v_\text{esc}$, yielding an interaction time
\begin{equation}
t_{int} \equiv \frac{R_p}{v_{esc}} \approx \sqrt{\frac{R_p^3}{\mathbf{2} G M_p}},
\end{equation}
where $M_p$ is the mass of the planet, and $G$ is the gravitational constant.
Taking the orbital period $P_{orb} \sim (a^3 / G M_\star)^{1/2}$, where $M_\star$ is the stellar mass and $a$ the orbit's semimajor axis, we have

\begin{equation}
\frac{t_{int}}{P_{orb}} \sim \mathbf{4\times10^{-3}} \Bigg(\frac{R_p}{R_{\earth}}\Bigg)^{3/2}\Bigg(\frac{0.1 \text{AU}}{a}\Bigg)^{3/2}\Bigg(\frac{M_{\star}}{M_{\odot}}\Bigg)^{1/2}\Bigg(\frac{M_{\earth}}{M_p}\Bigg)^{1/2},
\label{eq:interactiontimescale}
\end{equation}
with $R_\oplus$ and $M_\oplus$ the radius and mass of the Earth, respectively, and $M_\odot$ the solar mass.
The fact that this quantity is always $\ll 1$ implies that the interaction is effectively instantaneous relative to the orbital motion, and that it can therefore be modeled as an impulse to the planetary velocity at a fixed position, greatly simplifying the dynamics.

\subsection{Change in Period}

We are ultimately interested in the change in orbital period of the planet.
For small changes in period $\Delta P$, we have through Kepler's 3rd law that
\begin{equation}
\Bigg|\frac{\Delta P}{P}\Bigg| \approx \frac{3}{2}\Bigg|\frac{\Delta a}{a}\Bigg| \approx \frac{3}{2}\Bigg|\frac{\Delta E}{E}\Bigg|, \label{eq:kep3}
\end{equation}
where $\Delta a$ and $\Delta E$ are the corresponding changes in the semimajor axis and specific orbital energy, respectively, and in the second equality we used that $E = \frac{GM_\star}{2a}$.
In the impulse approximation, we can also easily relate the change in the specific energy to the change in the planet's velocity $V_p$, since at fixed planetary position, the only change in the planet's orbital energy is the (specific) kinetic energy $\frac{1}{2}V_p^2$,
\begin{equation}
\Delta E \approx \Delta \frac{1}{2} ({\bf V_p} \cdot {\bf V_p}) = ({\bf V_p} \cdot {\bf \Delta V_p}) + \mathcal{O}(V_p^2), \label{eq:dvp}
\end{equation}
where ${\bf V_p}$ and $\Delta {\bf V_p}$ are the planet's velocity and change in velocity, respectively, and bolded variables denote vector quantities. The change in energy is purely kinetic, so it only depends on the component of $\bf{\Delta V_p}$ parallel to the direction of orbital motion ${\bf V_p}/V_p$ (approximately azimuthal for nearly circular orbits). 
Defining the change in velocity along the orbit as $\Delta V_\phi \equiv ({\bf V_p} \cdot {\bf \Delta V_p})/ V_p$, we can combine Eqs.\:\ref{eq:kep3} and \ref{eq:dvp} to obtain the fractional change in the orbital period
\begin{equation}
    \frac{\Delta P}{P} \approx 3\frac{\Delta V_\phi}{V_p}.
    \label{eq:pShiftSimplified}
\end{equation}

\subsection{Change in Velocity}

Consider a parcel of gas with mass $\delta m$ leaving the planet at the escape velocity (Sec.\:\ref{sec:impulse}).
In the impulse approximation, the influence of the star is negligible, so we can simplify the analysis by considering only the planet and the parcel of gas, with their center of mass (COM) moving at a constant velocity $\bf{V_\text{COM}}$. 
The planet's inertial velocity is then given by $\bf{V_p} = \bf{V_\text{COM}} + \bf{V_P^\text{COM}}$, where $\bf{V_P^\text{COM}}$ is the planet's velocity in the center-of-mass frame. 
Because $\bf{V_\text{COM}}$ remains constant, the change in the planet's velocity is simply the change of the planet's center-of-mass velocity $\bf{V_P^\text{COM}}$.

In the COM frame  (Fig.\:\ref{fig:thetaDiagram}), the initial momentum of the planet must balance that of the escaping parcel (which we assume is launched at $V_\text{esc}$), so we have that the planet's initial velocity is
\begin{equation}
V_p^\text{COM} = \frac{\delta m}{M_p}{V_{\text{esc}}}.
\end{equation}

Since the parcel leaves at the escape velocity, it loses all its momentum at large separations.
To keep the COM stationary, the planet must also come to rest at late times.
The planet thus loses all its initial velocity in the COM frame, so $\Delta V = V_p^{\text{COM}}$ in the direction of the escaping wind.
The component $\Delta V_\phi$ in the direction of orbital motion (upward in Fig.\:\ref{fig:thetaDiagram}) is thus given by
\begin{equation}
    \Delta V_\phi = -\frac{\delta m}{M_p}{V_{\text{esc}}}\sin \theta. \label{eq:dvphi}
\end{equation}

The angle $\theta$ arises from the planet's orbital velocity, causing the approximately radial stellar wind \citep[e.g.,][]{Venzmer18} to have a head-on component in the planet's frame, like rain on the windshield of a moving bus \citep[Fig.\:\ref{fig:thetaDiagram}, see also Fig 1 of][]{Wang21b}. We have
\begin{equation}
\sin \theta = \frac{V_p}{\sqrt{V_p^2 + V_\text{wind}^2}} \approx {\frac{V_p}{V_\text{wind}}}, \label{eq:theta}
\end{equation}
where $V_\text{wind}$ is the stellar wind's velocity (assumed radial), both speeds are measured in the inertial frame, and the approximation holds in the limit $V_p \ll V_\text{wind}$. 
For a stellar wind speed of $250$ km/s \citep{Venzmer18}, this approximation holds to within $1\%$ for a 30-day orbit around a solar-mass star, but overestimates $\sin \theta$ by $\approx 30\%$ for ultra-short period planets with 1-day orbits. For such close-in planets (or slower stellar wind speeds), all fractional period shift equations in this paper can be corrected by the factor of $(1 + V_p^2/V_\text{wind}^2)^{-1/2}$ omitted in the approximation of Eq.\:\ref{eq:theta}.

Combining Eqs.\:\ref{eq:pShiftSimplified}, \ref{eq:dvphi} and \ref{eq:theta}, we therefore have that the escaping parcel of gas leaving from the surface of the planet causes a fractional shift in the planet's period given by
\begin{equation}
\frac{\Delta P}{P} = -3\frac{\delta m}{M_p} \frac{V_{\text{esc}}}{V_{\text{wind}}}.
    \label{eq:pShiftdm}
\end{equation}

In the limit of very fast winds such that $V_{\text{wind}} \gg V_p$, our expression predicts no change in period because the escaping gas will pull on the planet radially, which has no impact on the orbital period. In the opposite limit, where the stellar wind disappears and $V_{\text{wind}}$ approaches zero, the expected change in period seems to diverge, but this limit is unphysical. Our model only considers a stellar wind is strong enough to confine the escaping atmosphere into a tail. Assuming this is the case, a slower wind speed yields a more head-on geometry (Fig.\:\ref{fig:diagram}) and larger period shift.

\subsection{Tail Morphology}

\begin{figure}
\centering
    \includegraphics[width=\linewidth]{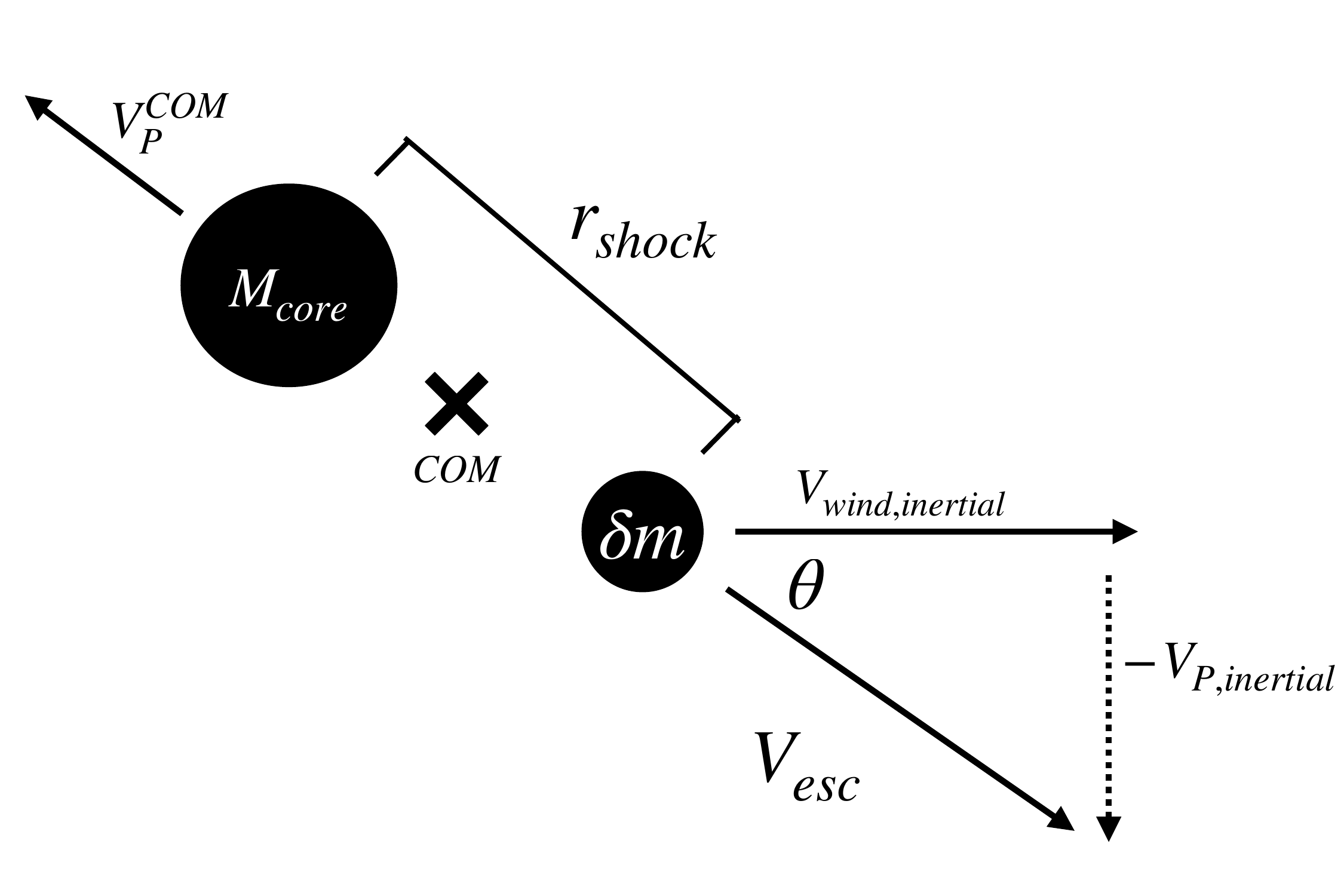}
    \caption{Initial conditions for the planet (large circle) and atmosphere (small circle), launched at escape velocity. At large separations both bodies come to rest, so the planet's total change in velocity is simply given by its initial velocity in the COM frame. The angle $\theta$ that the solar wind makes to the radial direction is due to the planet's orbital motion, giving the solar wind a head-on component in the COM frame.}
    \label{fig:thetaDiagram}
\end{figure}

The preceding section was ambiguous about the height from which the atmospheric parcel of gas is launched.
For our purposes, the key parameter is the height at which the escaping atmosphere transitions from being isotropically lost (with no effect on the orbital period) to being funneled into a localized tail (when it causes the planet period to decrease).
In our simplified model, we approximate this as the distance $r_\text{shock}$ from the center of the planet to the bow shock where the stellar wind confines and redirects the escaping flow.
We thus imagine the atmospheric parcel of gas in Fig.\:\ref{fig:thetaDiagram} as being launched at the escape velocity corresponding to $r_\text{shock}$.
Because the escape velocity falls off with distance from the planet as $r^{-1/2}$, particles launched from larger $r_\text{shock}$ have less momentum to take from the planet and a correspondingly smaller effect on the planet's orbital period.

We leave this distance $r_\text{shock}$ as a free parameter, as its calculation requires detailed hydrodynamic simulations and depends on the assumed heating and cooling processes, stellar wind properties, etc.
For example, in the 5 $M_\oplus$ evaporating planet simulations of \cite{Wang18}, $r_\text{shock} \approx 4-8 {R_{\earth}}$. With these modifications, equation \ref{eq:pShiftdm} becomes 

\begin{equation}
\frac{\Delta P}{P} = -3\frac{\delta m}{M_p} \frac{V_{\text{esc}}(R_\text{core})}
{V_{\text{wind}}} \left(\frac{r_{\text{shock}}}{R_{\text{core}}}\right) ^ {-1/2},
    \label{eq:pShiftrshock}
\end{equation}
where $V_{\text{esc}}(R_\text{core})$ is the escape velocity from the surface of the planet's core.

\subsection{Total Atmospheric Stripping}

Equation \ref{eq:pShiftrshock} is general, and one could calculate a cumulative change in orbital period by integrating over all the atmospheric gas parcels, keeping track of how the various quantities change with time. 

We now consider the limiting case where the atmosphere is stripped completely, as expected for small planets on short-period orbits are expected to lose their atmospheres entirely, either through photoevaporation \citep{Owen17} or core-powered mass loss \citep{GSS16}.
In the photoevaporation models of \cite{Owen17}, total atmospheric stripping occurs for planets with core radii smaller than $R_c^\text{crit}$, where 
\begin{equation}
    \frac{R_{c}^{\text{crit}}}{R_{\oplus}} \approx \Bigg(\frac{P}{30 \text{ days}}\Bigg)^{-1/4}.
    \label{eq:maxRadiusStripping}
\end{equation}

If we assume that the atmosphere makes up a small fraction of the planet's mass (so that the escape velocity remains approximately constant as the planet loses its atmosphere), and approximate the stellar wind properties and $r_\text{shock}$ as constant over the period of mass loss, total stripping of the atmosphere yields  

\begin{equation}
\frac{\Delta P}{P} = -3\frac{M_{\text{atm}}}{M_p} \frac{V_{\text{esc}}(R_\text{core})}
{V_{\text{wind}}} \left(\frac{r_{\text{shock}}}{R_{\text{core}}}\right) ^ {-1/2},
    \label{eq:pShiftMatm}
\end{equation}
where $M_\text{atm}/M_\text{core}\approx M_\text{atm}/M_p$ is  the initial atmospheric mass fraction.

\subsection{Stellar Wind} \label{sec:swind}
Typical stellar wind velocities are uncertain, given that they rely on observations from our own Sun, as well as magneto-hydrodynamic models of stellar atmospheres.  
The velocity profile typically rises as the wind is accelerated through magnetic fields and thermal pressure, asymptoting to a maximum velocity around $\sim 0.2$ AU \citep{Johnstone2015}. We use a reference wind speed of $V_{\text{wind}} = 250 \text{ km/s}$ adopted by \cite{Venzmer18}, and vary the wind speed between values used by \cite{McCann19} of 200 km/s and by \cite{Kislyakova14} of 400 km/s.

The stellar wind in our approximation is based on the "Strong Wind" case outlined in \cite{McCann19}, implying that $r_\text{shock}$ as described in section 2.4 remains inside the Hill Sphere of the planet, and it ignores interactions between atmospheric particles.
\subsection{Mass-Radius Relation}\label{sec:massradius}

Finally, we adopt the mass-radius relation for the planets' rocky cores from \cite{Valencia06},
\begin{equation}
    \frac{M_{\text{core}}}{M_{\oplus}} = \left(\frac{R_{\text{core}}}{R_{\oplus}}\right) ^4,
    \label{eq:mrRelation}
\end{equation}
which allows us to express the escape velocity in Eq.\:\ref{eq:pShiftMatm} only in terms of the core radius.

\subsection{Numerical Tests} \label{sec:numtests}

Although the hydrodynamics and radiative transfer involved in the interactions between the stellar wind and escaping atmosphere are complex and require detailed numerical simulations like those found in \cite{McCann19}, we can test our impulse approximation (\ref{eq:pShiftMatm}) through simple N-body gravitational simulations. In our numerical tests, we simulate the interaction of three particles, representing the star, planet, and a parcel of ejected atmosphere. The simulations initialize the planet at an orbital period of 25 days.
Because Eq.\:\ref{eq:pShiftMatm} assumes that parameters remain fixed over the period of mass loss, we model the lost atmosphere as a single particle with all the mass $M_{\text{atm}}$, launched from $r_{\text{shock}}=kR_\text{core}$ at its corresponding escape velocity\footnote{It is important to correct the traditional (test-particle) escape velocity expression to account for the significant atmospheric mass lost, $V_{\text{esc}} = \sqrt{\frac{2GM_{P}}{R_{P}}}\sqrt{1+M_{\text{atm}}/M_{\text{core}}}$.} and appropriate angle $\theta$ (Eq.\:\ref{eq:theta}). We also confirmed that splitting the atmosphere into many smaller particles did not affect the results.

\begin{figure}
    \centering
    \includegraphics[width=\linewidth]{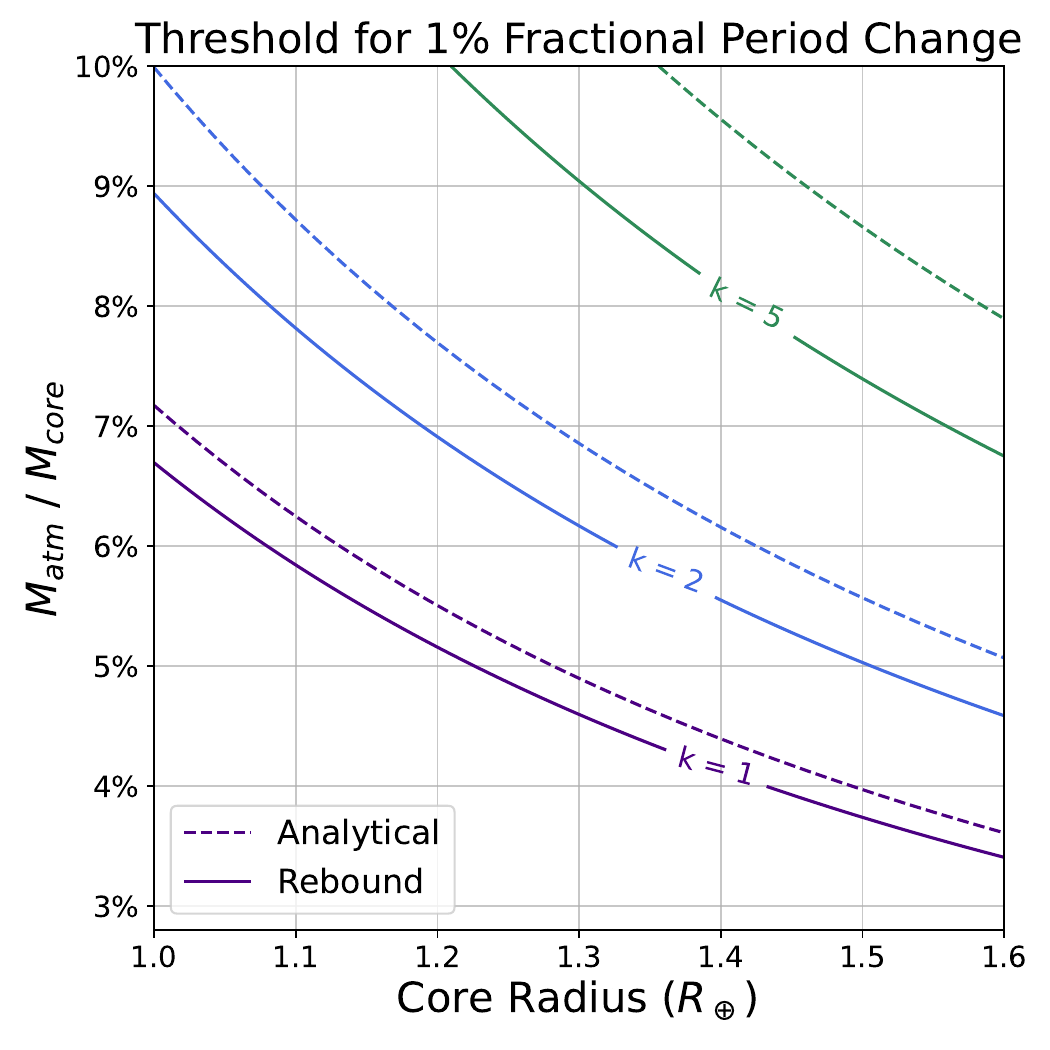}
    \caption{Contours for which the planet's orbital period decreases by 1\% from N-body simulations (solid) and the analytical expression in Eq.\:\ref{eq:pShiftMatm} (dashed), varying the height $k\equiv r_\text{shock}/R_\text{core}$ at which the stellar wind funnels the escaping atmosphere into a tail. All simulations fix the stellar wind velocity at 250 km/s, and the initial period at 25 days.}
    \label{fig:ContourPeriodShift}
\end{figure}

In Fig.\:\ref{fig:ContourPeriodShift}, we vary both the initial atmospheric mass fraction and the core radius, plotting the contours at which the orbital period shrinks by 1\% for different choices of $k\equiv r_\text{shock}/R_\text{core}$.

We used the adaptive-timestep IAS15 integrator \citep{Rein15} in the REBOUND N-body package \citep{Rein12}. 
We assume that after the short period of interaction (Sec.\:\ref{sec:impulse}), the continued action of the stellar winds over the much longer orbital period prevents the atmosphere from re-impacting the planet, so we halt the simulation after one orbital period.
Solid lines in Fig.\:\ref{fig:ContourPeriodShift} for the N-body results approximately match the simple expression in Eq.\:\ref{eq:pShiftMatm}. The disagreement between analytical and numerical predictions comes from the breakdown of the impulse approximation. As the atmosphere is formed into a tail at larger planetocentric distances, the interaction timescale (Eq.\ref{eq:interactiontimescale}) gets longer compared to the planet's orbital period, and the unmodeled effects from the star's tidal gravity become more important. 

\subsection{Initial Atmosphere Mass Fraction} \label{sec:initatm}

The fact that the planet's initial atmospheric mass fraction $M_\text{atm}/M_\text{core}$ enters linearly in Eq.\:\ref{eq:pShiftMatm} renders it an important parameter.
Initial atmospheres accreted during the protoplanetary disk phase are likely significant \citep{Lee15, Lee18, Fung18}.
However, as the protoplanetary disk disperses, removing the disk's pressure support on the planetary atmosphere, heavy mass loss is expected in a rapid ``boil-off" phase \citep{Owen16, GSS16}.
This phase could also lead to migration and merits its own modeling. However, it is not clear whether the stellar wind would be able to funnel the escaping atmosphere into a tail during this phase where the planet is embedded in the protoplanetary disk, so we do not consider it.

Analytically modeling this ``boil-off" phase, \cite{GSS16} derive a resulting atmospheric mass fraction post disk-dispersal of
\begin{equation}
    \frac{M_{\text{atm}}}{M_{\text{core}}} \approx A\Bigg(\frac{M_{c}}{M_{\oplus}}\Bigg)^{1/2},
    \label{eq:GSS16relation}
\end{equation}
with a constant of proportionality of $A\approx 0.01$.
\cite{Rogers23} started instead with observations of present-day planets with well measured radii and masses, and used the photoevaporation models of \cite{Owen17} to infer their initial atmospheres.
The best-fit relation to their results yielded a relation consistent with \cite{GSS16}.

By contrast, \cite{GSS18} needed a higher initial atmospheric mass fraction ($A \approx 0.05$) to match the observed radius valley when modeling the escape through core-powered mass loss.
We therefore adopt a fiducial value of $A=0.03$, intermediate between these results.

\section{Results} \label{sec:results}

Using both the mass-radius relation (\ref{eq:mrRelation}) and the initial atmosphere mass fraction (\ref{eq:GSS16relation}), we may substitute all masses in expression (\ref{eq:pShiftMatm}) for planet radii. We obtain an expression for the fractional period change under total atmospheric stripping

\begin{multline}
    \frac{\Delta P}{P} = -0.5\% 
    \left(\frac{R}{1.3 R_{\oplus}} \right)^ {7/2}
    \left(\frac{V_{\text{wind}}}{250 \text{km/s}} \right) ^{-1} \left(\frac{r_{\text{shock}}/{R_{\text{core}}}}{5}\right) ^ {-1/2}
    \label{eq:dShiftK2}
\end{multline}
where the reference value of $1.3 R_\oplus$ corresponds approximately to the center of the peak interpreted as stripped cores in the observed radius distribution \citep{Fulton17, Owen17, GSS18}, and the proportionality constant would vary between -0.2\% and -0.8\% depending on the values of $A$ discussed in Sec.\:\ref{sec:initatm}.

\begin{figure}
    \centering
    \includegraphics[width=1\linewidth]{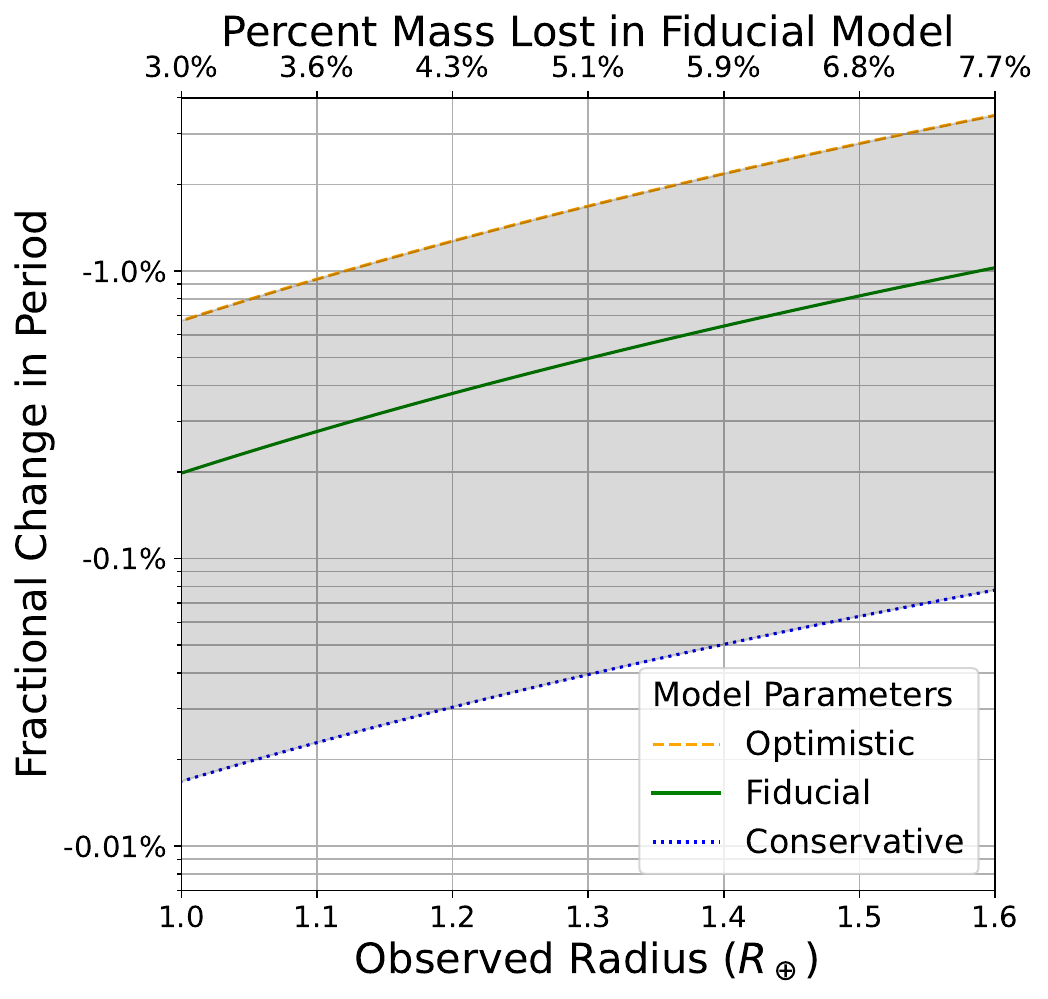}
    \caption{Fractional change in orbital period vs. observed radius assuming total atmospheric stripping, considering fiducial parameter values in Eq:\:\ref{eq:dShiftK2}, as well as optimistic and pessimistic cases (see text for adopted values). The fiducial model, in green, has $A=0.03$ (Eq.\:\ref{eq:GSS16relation}), $V_{\text{wind}}=250\text{km/s}, r_{\text{shock}} = 5 R_{\text{core}}.$}. 
    \label{fig:DiffMassRelations}
\end{figure}

We plot the above relation using our fiducial reference values as the green line in Figure \ref{fig:DiffMassRelations}. 
The relevant pileup interpreted as stripped cores in the observed radius distribution extends out to the right of the plot at $R\approx 1.6 R_\oplus$ \citep{Fulton17}.

We also plot a 'conservative' and an 'optimistic' model, which give lower and upper bounds for orbital migration. The conservative model uses values of $A=0.01$, $V_\text{wind}=400\text{km/s}$ and $r_\text{shock}/R_\text{core} = 10$. The optimistic model uses parameters $A=0.05$, $V_\text{wind}=200\text{km/s}$ and $r_\text{shock}/R_\text{core} = 1$. The broad range of expected period shifts is colored in gray. Because the conservative and optimistic models assume somewhat extreme values for all parameters simultaneously, we expect reasonable estimates to cluster closer to the fiducial green line than to the edges of the gray region.

We note that the fractional change in orbital period is a steep function of the core radius (i.e., the radius we observe today after stripping), varying from a few tenths of a percent to 1\% in our fiducial model across a narrow range of planet radii.

\subsection{Observable Consequences} \label{sec:jumps}

Although the anticipated fractional change in period given by our expression is modest, such inward migration may result in observable signatures for planet pairs with orbital periods near mean motion resonances (MMRs) at integer ratios. 
Given the expectation that inner planets will migrate inward more than their outer neighbors (since the outer neighbors are more likely to retain their atmospheres), the ratio of the pair's orbital periods will grow. When encountering MMRs under such divergent migration, capture into resonance is impossible, and the pair is forced to jump over the MMR, in the process receiving a kick to their orbital eccentricities \citep{Murray99}. This can be tested using precisely measured eccentricities using Transit Timing Variations (TTVs) for near-resonant planets \citep[e.g.,][]{Hadden17}. We explore this idea in detail in an upcoming paper. This jump can be observed in the distribution of exoplanets' distance from resonance.

\begin{figure}
    \centering
    \includegraphics[width=1\linewidth]{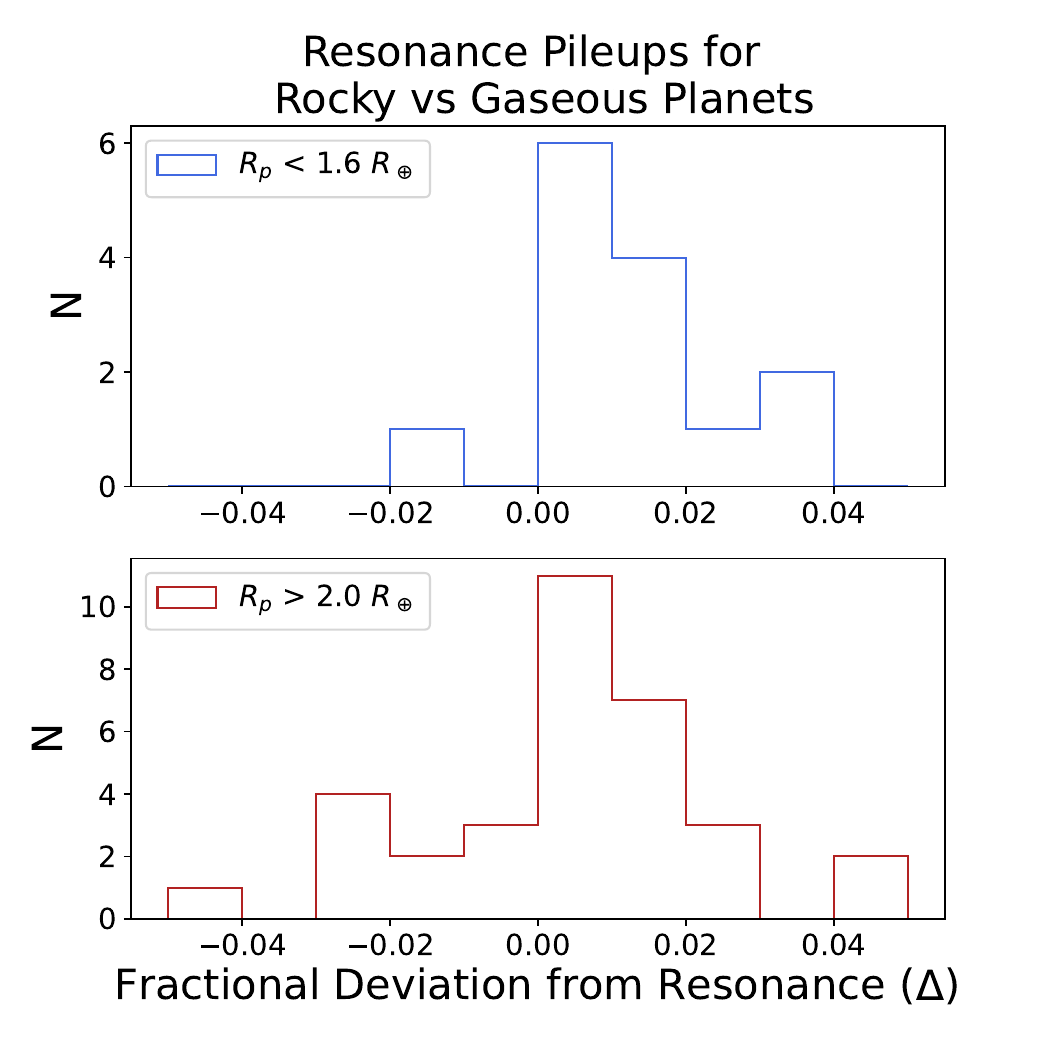}
    \caption{Histograms of the period ratio deviations from resonance $\Delta$ (Eq.\:\ref{eq:delta}) for a sample of 57 near-resonance exoplanet pairs from \cite{Hadden17}. The upper plot shows the distribution of deltas for rocky ($R_{p} < 1.6 R_{\oplus}$) planets (n=14). The lower plot shows the distribution for gaseous ($R_{p} > 2.0 R_{\oplus}$) planets (n=33). Deltas are calculated for each pair as the fractional deviation from the nearest mean motion resonance (MMR).}
    \label{fig:rocky_vs_gas}
\end{figure}

For a pair of planets with period ratio $P_\text{ratio}$, the fractional deviation $\Delta$ from the nearest MMR is given by \citep{Lithwick12}

\begin{equation}
\Delta = \frac{P_\text{ratio} - P_\text{res}}{P_\text{res}}, \label{eq:delta}
\end{equation}
where $P_\text{res}$ is the particular resonant period ratio. Figure \ref{fig:rocky_vs_gas} shows the distribution of $\Delta$ for two populations from the catalog of near-resonant pairs of \cite{Hadden17}, with planet radii taken from the NASA Exoplanet Archive. The top population consists of exoplanet pairs whose innermost planets are rocky exoplanets with observed radii $<1.6 R_\oplus$, while the bottom population represents those inner members are gaseous planets with radii $>2.0 R_\oplus$. The rocky cores have a more pronounced pileup wide of resonance, with fewer pairs with $\Delta < 0$. This is consistent with a picture in which smaller planets migrate inward as they lose their atmosphere to become rocky cores, thereby jumping over MMRs, while larger gaseous planets mostly maintain their atmospheres and remain in place. 
However, the statistics are marginal and more observations of near-resonant planets are needed to obtain a clearer picture.

Additionally, our model predicts that among the rocky cores $< 1.6 R_\oplus$, migration through atmospheric mass loss should increase with the planet radius. While the statistics degrade even further for this subgroup, we do not find any clear trends with planet radius for small planets. One possible interpretation is, since our mechanism depends on the uncertain initial masses of these planets' primordial atmospheres, that our assumed scaling of the initial atmospheric mass with planet radius (Eqs. \ref{eq:pShiftMatm} \& \ref{eq:GSS16relation}) may be  inaccurate.

\section{Conclusion} \label{sec:conclusion}

Drawing on hydrodynamic simulations of exoplanet atmosphere stripping, we focus on the case in which strong stellar winds confine the escaping atmosphere into a tail. We develop a kinematic framework to investigate the consequences of this kind of mass loss on the planetary orbit. 
We derive a simple expression (Eq.\:\ref{eq:maxRadiusStripping}) depending on observable and free parameters. We compare our expression with N-body integrations in \texttt{REBOUND} and find reasonable agreement over the relevant parameter space. 
Then, focusing on total atmospheric stripping, we calculate the expected migration (Eq.\:\ref{eq:dShiftK2}) using previous models for the initial atmospheric mass fraction. 
Over the relevant parameter space, we find that the period can decrease by a few tenths of a percent to a few percent. 

While this mechanism produces limited migration, it moves close-in planets inwards, causing divergent migration with any outer neighbor. This divergent migration separates our work from  previous literature \citep{Teyssandier15, boue12, Fujita22} which predict outward migration. For planets near MMRs, divergent migration will cause the pair to jump over the resonance, amplifying the extent of apparent migration and giving a kick to their orbital eccentricities \citep{Murray99}. By contrast, an outwardly migrating inner planet \citep{Teyssandier15, boue12, Fujita22} would lead to convergent migration and capture into resonance \citep{Murray_Dermott_2000}. Further study of how the radius valley maps onto the distribution of near-resonant planets could provide valuable independent observational constraints on migration driven by atmospheric mass loss. The model presented here could help clarify whether stellar winds are typically strong enough to funnel escaping atmospheres into planet-trailing tails as observed in simulations \citep{McCann19, Wang18}.
\section{Acknowledgments} 
This work was made possible in part through a gift by Xinyi Guo (Pomona ’12) to the Harvey Mudd Physics Summer Research Fund. The presented numerical calculations were made possible by computational resources provided through an endowment by the Albrecht family. This research has also made use of the NASA Exoplanet Archive, which is operated by the California Institute of Technology, under contract with the National Aeronautics and Space Administration under the Exoplanet Exploration Program.
\bibliography{Bib}{}
\bibliographystyle{aasjournal}
\end{document}